\documentclass[11pt,preprint,aps,showkeys,onecolumn]{revtex4}
\usepackage{mathrsfs}
\usepackage{amsmath}
\usepackage{amssymb}
\usepackage{amsfonts}
\usepackage{tikz}
\usepackage{hyperref}
\usepackage{xcolor}
\usepackage{epstopdf}
\usepackage{dcolumn}
\usepackage{bm}
\usepackage{natbib}
\usepackage{float}
\usepackage{afterpage}
\usepackage{caption}
\usepackage{subcaption}  
\usetikzlibrary{decorations.pathmorphing}
\usepackage[section]{placeins}
\newcommand{\be}{\begin{equation}}
\newcommand{\en}{\end{equation}}
 \newcommand{\bea}{\begin{eqnarray}}
 \newcommand{\ena}{\end{eqnarray}}

\begin{document}

\title{Matching McVittie spacetimes}
\author{Jining Tang$^{1}$}
\email{aishiker1998@gmail.com}
\author{Yang Huang$^{2}$}
\email{sps\_Huangy@ujn.edu.cn}
\author{Hongsheng Zhang$^{2}$}
\email{sps\_zhanghs$@$ujn.edu.cn}
\affiliation{School of Physics and Technology, University of Jinan,
336 West Road of Nan Xinzhuang, Jinan, Shandong 250022, China}

\begin{abstract}
	Gravitational collapse and bubble evolution in the asymptotic Friedmann-Lemaitre-Robertson-Walker (FLRW) Universe is an intriguing and intricate problem. We systematically analyze the dynamics of contact Schwarzschild-FLRW (McVittie) spacetimes, focusing on their general junction conditions and introducing a novel function to simplify the extrinsic curvature and surface stress-energy tensor. Both static and dynamic scenarios are explored, including special cases such as Schwarzschild, FLRW, and Einstein-Straus configurations using our general framework. Numerical calculations further investigate the evolution of concentric McVittie spacetimes under various initial conditions, incorporating $\Lambda$-CDM cosmological models to better reflect realistic cosmic backgrounds. These results offer deep insights into the interplay between the McVittie mass parameter, initial peculiar velocity, and the influence of dark energy, providing a unified perspective for understanding gravitational collapse and bubble evolution in cosmology and astrophysics.
\end{abstract}
\maketitle	
\section{Introduction}
\label{intro}

Since Oppenheimer and Snyder’s foundational study on gravitational collapse\cite{oppenheimer1939continued},  the dynamics of a spherically collapsing star with various forms of matter have been examined in detail \cite{caiBlackHoleFormation2006,chakrabortyCollapseDynamicsStar2010,gautreauGravitationalCollapseSingle1995,ilhaDimensionallyContinuedOppenheimer1999,joshiGravitationalCollapseStory2000,joshiRECENTDEVELOPMENTSGRAVITATIONAL2011,lakeCollapseRadiatingFluid1981,lakeCollapseRadiatingImperfect1982,shojaiGeneralizedOppenheimerSnyderGravitational2022,tippettGravitationalCollapseQuantum2011,lewandowskiQuantumOppenheimerSnyderSwiss2023,nakaoOppenheimerSnyderSpacetimeCosmological1992,thorne2000gravitation}. However, unlike Oppenheimer’s simplified model, current observations show that the universe is expanding\cite{hubble1929relation}, implying that the asymptotic background beyond the star is more appropriately described by the FLRW metric than by the Schwarzschild solution. Investigating how the asymptotic behavior of spacetime impacts the processes of structure formation and local collapse remains a crucial and intriguing question in the fields of gravitation and cosmology.

Over the last century, the concept of a cosmological black hole has been a widely studied topic, often considered as the final state of a collapsing star embedded within the expanding universe\cite{faraoni2019cosmological,firouzjaeeAsymptoticallyFRWBlack2010,firouzjaeeSPHERICALSYMMETRYBLACK2012,gaurBlackHolesEmbedded2023}. The most prominent solution is the McVittie solution, which was first introduced in the 1930s as a non-uniform, spherically symmetric distribution of mass around a point in an expanding universe\cite{mcvittie1933mass}.After decades of discussion, it has been established that McVittie’s solution can be interpreted as describing black holes within a spatially flat Friedmann-Lemaître-Robertson-Walker (FLRW) universe, though debates on its exact implications continue\cite{faraoniEvolvingBlackHole2013,kaloperMcVittiesLegacyBlack2010,lakeMoreMcVittiesLegacy2011,nolanPointMassIsotropic1998}. Other solutions also address inhomogeneities in a homogeneous universe, but they differ from the McVittie solution \cite{bonnorLocalDynamicsExpansion2000,carreraInfluenceGlobalCosmological2010,sultanaCosmologicalBlackHoles2005,gaoBlackHolesUniverse2011,faraoniEmbeddingBlackHoles2018}.  Since the McVittie metric describes a comoving black hole in an FLRW universe, one might wonder whether a process of gravitational collapse occurs within such a cosmological setting. This question has been explored by various researchers, who have proposed a range of solutions \cite{faraoniWhatFateBlack2009,gellerMassPatchFRW2018,sahaGravitationalCollapseMatter2023}.

It remains unclear whether a collapsing star can be considered as the source of the McVittie solution. Nolan was the first to show that a uniform-density star co-expanding with the universe could serve as a viable gravitational source for this solution\cite{nolanSourcesMcVittiesMass1993}.  Nandra et al. employed the tetrad-base method, formulated in the language of geometric algebra by Lasenby, Doran, \& Gull\cite{lasenbyGravityGaugeTheories1998}, to derive the McVittie solution in non-comoving coordinates\cite{nandraEffectExpandingUniverse2012}. They also applied the same method to construct a model of a uniformly dense spherical region embedded within a homogeneous universe, demonstrating that Nolan’s work can be recovered in non-comoving coordinates as a special case\cite{nandraDynamicsSphericalObject2013}. Using the junction condition, both Nolan and Nandra showed that a uniform-density interior can be smoothly matched to a McVittie exterior. However, the possibility of a collapsing interior being smoothly matched in a similar manner has seldom been discussed in recent literature.

The junction conditions play a crucial role in the study of gravitational collapse, with the Darmois-Israel junction condition and thin-shell formalism being the most commonly used\cite{darmois1927equations,israel1966singular,lakeRevisitingDarmoisLichnerowicz2017}. This formalism provides a convenient method for exploring the evolution of boundaries or shells in cosmology. Given the uncertainties surrounding the evolution of collapsing matter in McVittie spacetime, it is essential to undertake a comprehensive study and comparative analysis of similar processes, such as the formation of cosmic bubbles. The dynamics of these bubbles are generally complex, but by choosing appropriate matter contents for the regions, the problem can be simplified. Oppenheimer’s model, as mentioned earlier, represents the simplest case, while the concept of a false vacuum has also been proposed\cite{blauDynamicsFalsevacuumBubbles1987}. Berezin et al. systematically studied the evolution of thin-shell bubbles, developing a useful formalism based on Israel’s work\cite{berezinDynamicsBubblesGeneral1987}. Following this, the junction of arbitrary FLRW spacetimes has been studied in depth\cite{sakaiJunctionConditionsFriedmannRobertsonWalker1994}.  It is widely accepted that two FLRW spacetimes cannot be matched without a shell. The symmetry of the bubbles makes it possible to express the thin-shell formalism in a greatly simplified form \cite{lakeEquationMotionBubble1984,lakeThinSphericalShells1979}. On the other hand, the gluing of McVittie spacetimes, as an analogy to Oppenheimer’s model in cosmology, has received little attention due to its complex non-linear coupling\cite{haines1993thin}. Like other cosmological black hole solutions, McVittie’s solution can reduce to the Schwarzschild and FLRW forms by adjusting the appropriate parameters. Therefore, it is both sensible and essential to investigate the junction of two McVittie spacetimes and compare its asymptotic behavior with previously studied classical solutions.

In this paper, we propose a novel approach to address the junction of two McVittie-like regions and evaluate its implications. The structure of this paper is as follows.  In Sect.~\ref{sec:2}, we review the McVittie solution and Israel's thin shell formalism in the chosen coordinates; in Sect.~\ref{sec:3}, we construct a model of the McVittie bubble and calculate its relevant physical quantities. We then introduce a new function and explore its significance within the context of thin-shell formalism. Sect.~\ref{sec:4} presents several applications of our model, detailing special cases using the newly proposed function. Finally, Sect.~\ref{sec:5} provides our conclusions. In this work, we adopt the natural unit system where $c = G = 1$ and the metric signature $(- + + +)$.

\section{Review of McVittie metric}
\label{sec:2}
 For the convenience of developing the investigation in this paper, we briefly review the McVittie metric and the junction condition.
 
\subsection{McVittie's solution}
\label{sec:A}
The solution discovered by McVittie ninety years ago was originally formulated to describe a black hole embedded in an FLRW universe. This spherically symmetric solution is parameterized by the asymptotic scale factor $a(t)$ of the cosmology and a parameter  $m$, which represents the mass of the central black hole,
\begin{equation}
	ds^2=-\left(\frac{1-\frac{m}{2 a(t)\bar{r}}}{1+\frac{m}{2 a(t)\bar{r}}}\right)^2dt^2+a^2(t)\left(1+\frac{m}{2 a(t)\bar{r}}\right)^4\left(d\bar{r}^2+\bar{r}^2d\Omega^2\right).
\end{equation}

In McVittie’s original publication, the solution metric was presented in isotropic form, which allows the solution to reduce to both the Schwarzschild and FLRW solutions in the appropriate limits. The energy density and pressure of this solution can be derived by solving Einstein’s field equations, as shown in
\begin{equation}
	\rho(t)=\frac{3}{8\pi}H^2(t) \quad,\quad p=-\frac{1}{8\pi}\left[3H^2+2\dot{H}\frac{1+\frac{m}{2a(t) \bar{r}}}{1-\frac{m}{2a(t) \bar{r}}}\right],
\end{equation}
where $H\equiv H(t)=\frac{\dot{a}}{a}$ is the Hubble function of the FLRW background cosmology, and $m$ is a constant because of McVittie's no-accretion hypothesis($G^1{}_0=0$).  It is clear that the energy density is homogeneous for any value of $m$, but the second term of pressure is inhomogeneous and singular at $\bar{r}=\frac{m}{2a(t)}$ unless $\dot{H}$ vanishes.

The line element can be rewritten in terms of the areal radius $r$ as
\begin{equation}
	r\equiv a(t)\bar{r}\left(1+\frac{m}{2 a(t)\bar{r}}\right)^2,
\end{equation}
which offers a more intuitive physical interpretation. Under this transformation, the new metric is no longer diagonal, and is expressed as
\begin{equation}
	ds^2=-\left(1-\frac{2m}{r}-H^2r^2\right)dt^2-\frac{2Hr}{\sqrt{1-\frac{2m}{r}}}drdt+\frac{dr^2}{1-\frac{2m}{r}}+r^2d\Omega^2.
\end{equation}
When $m = 0$, the metric corresponds to that of a spatially flat FLRW universe in physical coordinates, and setting  $H = 0$  recovers the standard Schwarzschild metric. In this coordinate system, the energy density remains the same as previously calculated, while the pressure is given by
\begin{equation}
	p(t,r)=\rho(t)\left(\frac{1}{\sqrt{1-\frac{2m}{r}}}-1\right)+\frac{p_b}{\sqrt{1-\frac{2m}{r}}},
\end{equation}
where $p_b$ is the pressure of the background universe in the absence of mass.

Kaloper et al.’s review \cite{kaloperMcVittiesLegacyBlack2010} and a series of works by Faraoni \cite{faraoni2019cosmological, faraoniCosmologicalExpansionLocal2007, faraoniEmbeddingBlackHoles2018, faraoniEvolvingBlackHole2013} offer in-depth discussions on the apparent horizons and causal structure of the McVittie solution, which have contributed to a better understanding of this spacetime.

\subsection{Thin shell formalism}
\label{sec:B}

The thin shell formalism, developed by Israel, is based on the Gauss-Codazzi equations, which project the Einstein field equations onto a hypersurface \cite{babaevArtGluingSpaceTime}. The resulting junction conditions describe the relationship between the discontinuity in extrinsic curvature and the surface stress-energy tensor, providing a crucial framework for analyzing discontinuous spacetimes.

For a time-like hypersurface $\Sigma$ along with two glued spacetime manifolds $\mathcal{M}^\pm$, the parametric equation is
\begin{equation}
	f(x^\alpha(y^a))=0,
\end{equation}
where $x_\pm^\alpha$ is the coordinate system of $\mathcal{M}^\pm$ and $y^a$ is the coordinates on the hypersurface.
The unit normal vector in $M^\pm$ is given by the form
\begin{equation}
	n_\alpha= \pm\frac{1}{|g^{\mu\nu}\frac{\partial f}{\partial x^\mu}\frac{\partial f}{\partial x^\nu}|^{1/2}}\frac{\partial f}{\partial x^\alpha}\quad,\quad n_\alpha n^\alpha=1.
\end{equation}
Define the triad $e_a=\frac{\partial}{\partial y^a}$ tangent to the surface, whose components $e^\alpha _a=\frac{\partial x^\alpha}{\partial y^a}$, then the induced metric of $\mathcal{M}^\pm$ can be given by $g_{ab}=g_{\alpha\beta}e^\alpha _a e^\beta _b$.
The first junction condition for two gluing spacetimes proposed by Darmois requires
\begin{equation}
	[g_{ab}]_\pm \equiv g^+_{ab}-g^-_{ab}=0.
\end{equation}

The extrinsic curvature of $\Sigma$ is defined as
\begin{equation}
\begin{gathered}
	K_{ab}^\pm=e^\alpha _a e^\beta _b\nabla_\alpha n_\beta |_\pm
	=-n_\sigma\left[\frac{\partial^2 x^\sigma(y)}{\partial y^a \partial y^b}+\Gamma^\sigma _{\mu\nu}\frac{\partial x^\mu(y)}{\partial y^a}\frac{\partial x^\nu(y)}{\partial y^b}\right].
\end{gathered}
\end{equation}

The extrinsic curvature plays a fundamental role in relating the bulk spacetime to the hypersurface. According to the second junction condition, this geometric discontinuity is directly linked to the surface stress-energy tensor  $S_{ab}$, which encapsulates the physical properties of the thin shell,
\begin{equation}
	S_{ab}=-\frac{1}{8\pi}\left([K_{ab}]_\pm-[K]h_{ab}\right),
\end{equation}
where $K=K^a{}_a$ is the trace of the extrinsic curvature tensor.

The surface stress-energy tensor  $S_{ab}$  characterizes the physical properties of the hypersurface, with its trace representing the surface energy density, while its spatial components determine the stress or pressure within the shell. In a spherically symmetric system,  $S_{ab}$  can be decomposed into the surface energy density  $\sigma$  and the isotropic pressure $p$.

\section{Junction Conditions for Contacting McVittie Spacetimes}
\label{sec:3}
In this section, we investigate the junction conditions for contacting McVittie spacetimes, which resemble FLRW bubbles but are described in physical coordinates.
\subsection{The Bubble-like Model }
\label{sec:C}
In \cite{nandraDynamicsSphericalObject2013}, Nandra et al. introduced a model in which a spherical massive object with a uniform interior density $\rho_i(t)$ is embedded in an expanding universe with a uniform exterior density $\rho_e(t)$. They demonstrated that if the independent, spatially uniform Hubble parameters of the two regions, $H_i(t)$ and $H_e(t)$, are identical, the model coincides with the Nolan solution\cite{nolanSourcesMcVittiesMass1993}, which describes a uniform-density object within a spatially flat expanding universe. Additionally, we find that if the exterior spacetime is governed by the McVittie metric, the quantity 
 $(\rho_i(t) - \rho_e(t))a^3(t)$ must remain constant, where $a(t)$ defines the boundary of the spherical object.
In Nandra’s model, a smoothly matched condition is assumed, with a single coordinate system  $(t, r, \theta, \phi)$  applied uniformly across both the inner and outer regions. This approach simplifies the matching process by maintaining global time and radial coordinates. However, in general, such an assumption may not always hold, as different regions—especially those with distinct spacetime geometries or physical properties—typically require separate coordinate definitions for  $t$  and  $r$  to accurately describe their respective metrics. To address this limitation and ensure maximal generality, we introduce a new model, illustrated in Fig.1, in which both the interior and exterior regions are governed by the McVittie metric.

For the two regions denoted by $V^\pm$, the line elements take the form
\begin{equation}
	ds^2_\pm=-\left(1-\frac{2m_\pm}{r_\pm}-H_\pm(t_\pm)r^2_\pm\right)dt^2_\pm-\frac{2H_\pm(t_\pm)r_\pm}{\sqrt{1-\frac{2m_\pm}{r_\pm}}}dr_\pm dt_\pm+\frac{dr^2_\pm}{1-\frac{2m_\pm}{r_\pm}}+r^2_\pm d\Omega^2,
\end{equation}
where $\{x^\mu _\pm\} \equiv \{t_\pm,r_\pm,\theta,\phi\}$ represent two distinct coordinate systems. The spacetime structure is significantly influenced by variations in the parameters $m_\pm$ and $H_\pm(t_\pm)$.
Since $g^{rr} = 0$ leads to a time-dependent cubic equation, which generally admits three roots, appropriate parameter choices can ensure the existence of two positive real roots after the critical time $t_*$. These roots correspond to the event horizon of the central mass and the cosmological horizon of the background universe.\footnote{For dust-filled background($H(t)=\frac{2}{3t}$), when $t>t_*=2\sqrt{3}m$, have $m<\frac{1}{3\sqrt{3}H(t)}$ with two real horizons }.

\begin{figure}[htbp]
    \centering
    \begin{tikzpicture}

\draw[fill=gray!20, thick] (-3.5,-3.5) rectangle (3.5,3.5);

\node[above right] at (-2,2.7) {\footnotesize Background McVittie Spacetime};

\draw[fill=blue!20, thick] (0,0) circle (2.2cm);

\fill[black] (0,0) circle (3pt); 
\node[below] at (0,-0.2) {\scriptsize $m_-$}; 

\node at (-0.5,1) {\scriptsize $H_-$};

\draw[thick] (0,0) circle (2.2cm);

\node[below] at (0,-2.3) {\scriptsize Thin Shell};

\node[above] at (0,2.2) {\scriptsize $r=R(\tau)$};

\node[above right] at (1.4,2.2) {\scriptsize $m_+, H_+$};

\node at (-1,0) {\footnotesize Interior};

\draw[thick, red,<->, decorate, decoration={snake, amplitude=0.5mm, segment length=2mm}]
    (3,-1.5) -- (2,-1) ;
\draw[thick, red,<->, decorate, decoration={snake, amplitude=0.5mm, segment length=2mm}]
    (-3,1.5) -- (-2,1);

\draw[thick,red, <->, decorate, decoration={snake, amplitude=0.5mm, segment length=2mm}]
    (0.5,1.5) -- (0.2,0.5) ;
\draw[thick,red, <->, decorate, decoration={snake, amplitude=0.5mm, segment length=2mm}]
    (-0.5,-1.5) -- (-0.2,-0.5);

\end{tikzpicture}
    \caption{Schematic of the thin-shell model in McVittie spacetime (color version available online). The thin shell divides the interior McVittie region (blue/dark gray) from the exterior McVittie background (gray). Red wavy arrows represent the Hubble flow, illustrating the cosmological dynamics of the McVittie spacetime.}
    \label{fig:thin_shell}
\end{figure}

\subsection{Second fundamental form}
\label{sec:D}
Following the approach introduced in Sect.~\ref{sec:B}, the induced metric of the two regions must remain continuous across the hypersurface  $\Sigma$, which requires
\begin{equation}
	ds^2_\Sigma=ds^2_+|_\Sigma=ds^2_-|_\Sigma.
	\end{equation}
The parametric equations defining  $\Sigma$  in the two regions are given by
\begin{equation}
	f_\pm(t_\pm,r_\pm)=r_\pm-R_\pm(T_\pm(\tau))=0.
\end{equation}
Substituting these conditions, the metrics reduce to
\begin{equation}
	ds^2_\pm=-\left[\left(1-\frac{2m}{R}-H^2 R^2\right)\dot{T}^2+2\frac{HR}{\sqrt{1-\frac{2m}{R}}}\dot{T}\dot{R}-\frac{\dot{R}^2}{1-\frac{2m}{R}}\right]d\tau^2+R^2 d\Omega^2|_\pm.
\end{equation}

To simplify notation, subscripts will generally be omitted in intermediate steps to avoid unnecessary clutter. However, in key expressions where their distinction is crucial, subscripts will be explicitly retained to ensure clarity.
The 3D metric on the hypersurface is given by
\begin{equation}
	ds^2_\Sigma=-d\tau^2+\mathcal{R}^2(\tau)d\Omega^2.
\end{equation}

The first junction condition imposes the following constraint:
\begin{equation}
	\left(1-\frac{2m}{R}-H^2 R^2\right)\dot{T}^2+2\frac{HR}{\sqrt{1-\frac{2m}{R}}}\dot{T}\dot{R}-\frac{\dot{R}^2}{1-\frac{2m}{R}}=1 \quad,\quad R(T)=\mathcal{R}(\tau).
\end{equation}

The four-velocity tangent to the hypersurface and the corresponding four-acceleration in 4D coordinates are given by
\begin{equation}
	u^\alpha=\frac{\partial}{\partial \tau} =\dot{T}\frac{\partial}{\partial t}+\dot{R}\frac{\partial}{\partial r},
\end{equation}
and
\begin{equation}
		\begin{gathered}
			a^t=\frac{2 \dot{R} \dot{T} \left(m-R^3 H^2\right)}{R(R-2m)}+\frac{H \dot{R}^2}{\left(1-\frac{2 m}{R}\right)^{3/2}}
			-\frac{\dot{T}^2 H\left(m-R^3 H^2 \right)}{\sqrt{R (R-2 m)}}+\ddot{T},\\
			a^r=\dot{T}^2 \left(-\sqrt{R (R-2 m)} H'+H^2 (m-R)+R^3 H^4+\frac{m (R-2 m)}{R^3}\right)\\
			+\frac{2 H \dot{R} \dot{T} \left(m-R^3 H^2\right)}{\sqrt{R (R-2 m)}}-\frac{\dot{R}^2 \left(m-R^3 H^2\right)}{R(R-2m)}+\ddot{R},\\
			a^\theta=a^\phi=0.
		\end{gathered}
	\end{equation}

From Eq. (16), the unit outward normal $n_\alpha$ can be concisely expressed as
\begin{equation}
	n_\alpha=\left(-R(\tau),T(\tau),0,0\right).
\end{equation}

Substituting the previous expressions into the definition of the extrinsic curvature tensor (9), the non-vanishing components of  $K_{ab}$  are given by

\begin{equation}
K_{\tau\tau} = -n_\mu a^\mu = -n_t a^t - n_r a^r,
\end{equation}
which expands to
\begin{equation}
   \begin{aligned}
K_{\tau\tau} &= \dot{R} \ddot{T} - \dot{T} \ddot{R} + \frac{3m }{R(R - 2m)} \dot{T} \dot{R}^2 - \frac{m (R - 2m)}{R^3} \dot{T}^3 \\
&\quad + \left[\left(\frac{4m^2}{R} - 4m + R \right) \frac{H’}{\left(1 - \frac{2m}{R}\right)^{3/2}} + H^2 (R - m) - H^4 R^3 \right] \dot{T}^3 \\
&\quad - \frac{H^2 R^5 }{(R - 2m)} \dot{T} \dot{R}^2
+ \frac{3 (H^3 R^3 - mH)}{R\left(1 - \frac{2m}{R}\right)^{1/2}} \dot{R} \dot{T}^2
+ \frac{H\dot{R}^3}{\left(1 - \frac{2m}{R}\right)^{3/2}}.
\end{aligned} 
\end{equation}

Similarly, the angular component is

\begin{equation}
K_{\theta\theta} = -n_{\theta;\theta} = R \left[ \frac{H R \dot{R}}{\sqrt{1 - 2m/R}} + \left( 1 - \frac{2m}{R} - H^2 R^2 \right) \dot{T} \right].
\end{equation}

 Differentiating the first fundamental form (16) with respect to proper time yields
\begin{equation}
		\dot{\mathcal{R}}=\dot{a}\chi_0=\frac{dR}{dT}\frac{dT}{d\tau}=\dot{R},
	\end{equation}
and
\begin{equation}
		\begin{gathered}
			\frac{2 R^{2} \dot{R} \dot{T}^2 H'}{\sqrt{1-\frac{2m}{R}}}-2 R^2 HH' \dot{T}^3 -3 R H^2\dot{T}^2 \dot{R}+\frac{\dot{R}\dot{T}^2 }{R}
			+\frac{2 R^{2} H \ddot{R} \dot{T}}{\sqrt{1-\frac{2m}{R}}}\\+\frac{2 R^{2} H \dot{R} \ddot{T}}{\sqrt{1-\frac{2m}{R}}}-\frac{ H \dot{R}^2 \dot{T}}{(1-\frac{2m}{R})^{3/2}}
			+\frac{3  H \dot{R}^2 \dot{T}}{\sqrt{1-\frac{2m}{R}}}-\frac{\dot{R} \dot{T}^2 \left(1-\frac{2m}{R}-H^2R^2\right)}{R}\\
			+2 \dot{T} \ddot{T} \left(1-\frac{2m}{R}-H^2R^2\right)+\frac{R \dot{R}^3}{(R-2m)^2}-\frac{\dot{R}^3}{R-2m}-\frac{2 R \dot{R} \ddot{R}}{R-2m}=0.
		\end{gathered}
	\end{equation}
By substituting the above equation and the first fundamental form into  $K_{\tau\tau}$  to eliminate  $\ddot{R}$  and  $\dot{R}$, we obtain
\begin{equation}
		\begin{gathered}
		K^\tau{}_\tau=	H\sqrt{F \dot{T}^2-1}+\frac{H\dot{T}^2}{  \sqrt{F \dot{T}^2-1}}\frac{m}{R}+\frac{2 m}{R^2} \dot{T}+ \sqrt{\frac{F}{F\dot{T}^2-1}} \ddot{T} ,
		\end{gathered}
	\end{equation}
	where  $F \equiv F(R) = 1 - \frac{2m}{R} $.
It follows that when the Hubble parameter vanishes, only the last two terms remain.
	
	Now, we demonstrate that in the two asymptotic limits, the extrinsic curvature of the McVittie spacetime reduces to the well-established forms corresponding to the Schwarzschild and FLRW spacetimes.

Asymptotically Schwarzschild Case $( H \to 0 )$
	\begin{equation}
		K^\tau{}_\tau|_{\text{Schwarzschild}}= \frac{2 m}{R^2} \dot{T}+ \sqrt{\frac{F}{F\dot{T}^2-1}} \ddot{T}=\frac{\partial_\tau\left(F\dot{T}\right)}{\dot{R}}=\frac{\dot{\beta}}{\dot{R}},
	\end{equation}
	\begin{equation}
		K^\theta{}_\theta|_{\text{Schwarzschild}}=\frac{F\dot{T}}{R}=\frac{\beta}{R},
	\end{equation}
	where $\beta\equiv F\dot{T}=\sqrt{\dot{R}^2+F}$ is the well-known quantity introduced in \cite{poisson2004relativist}. This confirms that the extrinsic curvature correctly reduces to the Schwarzschild case in this limit.

Asymptotically Spatially Flat FLRW Universe $( m/R \to 0 )$
	\begin{equation}
		K^\tau{}_\tau|_{\text{FLRW}}= -H\sqrt{\dot{T}^2-1}-\frac{\ddot{T}}{\sqrt{\dot{T}^2-1}},
	\end{equation}
	\begin{equation}
		K^\theta{}_\theta|_{\text{FLRW}}=\frac{(1-H^2R^3)}{R}\dot{T}+H\dot{R},
	\end{equation}
	For a comoving shell, where $T(\tau)=\tau,R=S(\tau)r_0$, the above components simplify to
	\begin{equation}
		K^\tau{}_\tau|_{\text{FLRW}}=0\quad,\quad K^\theta{}_\theta|_{\text{FLRW}}=\frac{1}{S(\tau) r_0}.
	\end{equation}
This result is consistent with the standard expression for the extrinsic curvature in FLRW spacetime.

Solve the first equation in (16), we found the self-consistent relation of $\dot{T}$ and $\dot{R}$ is
\begin{equation}
	\left(1-\frac{2m}{R}-H^2R^2\right)\dot{T}=\sqrt{1-\frac{2m}{R}-H^2R^2+\dot{R}^2}-\frac{HR}{\sqrt{1-2m/R}}\dot{R}.
\end{equation}

As an analog of the function $F(R)$ and $\beta(R,\dot{R})$ in the Schwarzschild case, we define the new functions
\begin{equation}
	\mathcal{F}\equiv\mathcal{F}(T,R)=1-\frac{2M(T,R)}{R}=1-\frac{2m}{R}-H^2R^2,
\end{equation}
and
\begin{equation}
	\mathcal{B}\equiv\mathcal{B}(T,R,\dot{R})=\sqrt{\dot{R}^2+\mathcal{F}}=\mathcal{F}\dot{T}+\frac{HR}{\sqrt{1-2m/R}}\dot{R},
\end{equation}
where $M(T,R)$ is the gravitational mass of a spherical region in radius $R$ in McVittie spacetime.

By comparing (33) with (21) and (22), we find that the cumbersome expression can be reformulated more succinctly using the $\mathcal{B}$ function.  In particular, the angular component simplifies to the compact form
\begin{equation}
	K^\theta{}_\theta^+|_\textbf{McV}=\frac{1}{R}\left[\frac{H R \dot{R}}{\sqrt{1-2 m/R}}+\left(1-\frac{2m}{R}-H^2R^2\right)\dot{T}\right]=\frac{\mathcal{B}}{R},
\end{equation}
whereas the time component
\begin{equation}
		K^\tau{}_\tau^+|_\textbf{McV}=\frac{\dot{\mathcal{B}}}{\dot{R}}+\frac{ H'R \dot{T}}{\mathcal{B}\dot{R}}\left(HR-\sqrt{1-\frac{2m}{R}} \dot{R} \dot{T}\right),
	\end{equation}
still contains a nonlinear term arising from the dynamical background, whose interpretation and implications will be discussed in the next subsection.

\subsection{Thin shell quantities}
\label{sec:E}

In general, a shell exists at the junction between two spacetimes, representing a discontinuity or an interface. For a spherical shell composed of a perfect fluid, the surface stress-energy tensor $S_{ab}$ takes the form
\begin{equation}
	S_{ab}=(\sigma+\varpi)u_au_b+\varpi h_{ab},
\end{equation}
Applying the thin-shell formalism (10), the shell quantities can be directly obtained as
\begin{equation}
	\sigma=-\frac{1}{4\pi} \left[K^\theta{}_\theta\right]_\pm=-\frac{1}{4\pi }\left[\frac{\mathcal{B}}{R}\right]_\pm,
\end{equation}
and
\begin{equation}
	\begin{aligned}
		&\varpi=\frac{1}{8\pi}\left(\left[K^\theta{}_\theta\right]_\pm+\left[K^\tau{}_\tau\right]_\pm\right)\\&=\frac{1}{8\pi}\left(\left[\frac{\mathcal{B}}{R}\right]_\pm+\left[\frac{\dot{\mathcal{B}}}{\dot{R}}\right]_\pm+\left[\frac{ H'R \dot{T}}{\mathcal{B}\dot{R}}\left(HR-\sqrt{1-\frac{2m}{R}} \dot{R} \dot{T}\right)\right]_\pm\right).
	\end{aligned}
\end{equation}

By differentiating (37) with respect to $\tau$ and substituting the result into (38), the equation of state for the shell can be derived
\begin{equation}
	\dot{\sigma}+2\frac{\dot{R}}{R}\left(\sigma+\varpi\right)=\frac{1}{4\pi}\left[\frac{H'\dot{T}}{\mathcal{B}}\left(HR-\sqrt{1-\frac{2m}{R}} \dot{R} \dot{T}\right)\right]_\pm.
\end{equation}
The above equation is equivalent to
\begin{equation}
			\frac{d}{d\tau}(A_s\sigma)+\varpi\frac{dA_s}{d\tau}=A_s\left(\dot{\Phi}_+-\dot{\Phi}_-\right),
		\end{equation}
		where $A_s\equiv 4\pi R^2$ is the surface area of the shell, and
		\begin{equation}
			\dot{\Phi}=\frac{H'\dot{T}}{4\pi\mathcal{B}}\left(HR-\sqrt{1-\frac{2m}{R}} \dot{R} \dot{T}\right).
		\end{equation}
Equation (40) can be interpreted as follows: the first term on the left-hand side quantifies the change in the total energy of the shell, while the second term represents the work done by the surface pressure due to variations in the shell’s area. The term on the right-hand side accounts for the net flux rate across the shell.

For the line elements given in Eq. (11), the expression (33) can be rewritten as
\begin{equation}
		\mathcal{B}_\pm=\sqrt{\dot{R}^2+1-\frac{2m_\pm}{R}-H^2_\pm R^2}=\sqrt{\dot{R}^2+1-\frac{2m_\pm}{R}-\frac{8\pi\rho_\pm}{3} R^2},
	\end{equation}
where the Friedmann equation has been used.

The mass of the shell, derived from Eq. (36), is given by
\begin{equation}
	\begin{aligned}
			M_s&=4\pi R^2\sigma=R\left(\mathcal{B}_--\mathcal{B}_+\right)\\&=R\left(\sqrt{\dot{R}^2+1-\frac{2m_-}{R}-\frac{8\pi\rho_-}{3} R^2}-\sqrt{\dot{R}^2+1-\frac{2m_+}{R}-\frac{8\pi\rho_+}{3} R^2}\right).
\end{aligned}
		\end{equation}
Combining (37), (42) and (43), we obtain
\begin{equation}
	\mathcal{B}_\pm=\frac{\mathcal{B}_+^2-\mathcal{B}_-^2\pm(\mathcal{B}_+-\mathcal{B}_-)^2}{2(\mathcal{B}_+-\mathcal{B}_-)}=\frac{m_+-m_-}{4\pi R^2\sigma}+\frac{\rho_+-\rho_-\mp6\pi\sigma^2}{3\sigma}R.
\end{equation}

If the shell vanishes ($\sigma=0$), the relationship between parameter $m$ and $H$ must satisfy
\begin{equation}
		m_+-m_-=\frac{4\pi}{3}\left(\rho_--\rho_+\right)R^3=\frac{1}{2}(H_-^2-H_+^2)R^3.
	\end{equation}
 However, according to the definition of McVittie’s solution, which assumes the constancy of the mass parameter $m_\pm$, we need
\begin{equation}
	\frac{4\pi}{3}\left(\rho_--\rho_+\right)R^3=(H_-^2-H_+^2)R^3=\mathbf{Const.}
\end{equation}
to hold. Evidently, with appropriate parameter choices, two Kottler spacetimes can be smoothly matched. Moreover, the independent derivation of the Nolan solution in Nandra’s work further confirms the validity of the above equation\cite{nandraDynamicsSphericalObject2013}.

\section{Applications}
\label{sec:4}
Here, we present several applications by considering asymptotic cases of our model, obtained by selecting specific parameter choices that simplify the general junction of two McVittie spacetimes. These applications help to explore the distinctive features of the model and the conditions governing its evolution.

\subsection{Dust thin shell I: Minkowski interior}
\label{sec:F}

For the case consisting of pressureless dust, the stress-energy tensor is given by
	\begin{equation}
		S_{ab}=\sigma u_a u_b.
	\end{equation}
	
	We assume that the interior spacetime is flat,  while the exterior spacetime is static, i.e.,  $(dH/dT=0, H=H_0)$, implying the last term in (37) vanishes, then the thin-shell formalism yields
	\begin{equation}
		\begin{gathered}
			\sigma=\frac{1}{4\pi R}\left(\mathcal{B}_--\mathcal{B}_+\right)\quad, \quad\varpi=\frac{\mathcal{B}_+-\mathcal{B}_-}{8\pi R}+\frac{\dot{\mathcal{B}}_+-\dot{\mathcal{B}}_-}{8\pi \dot{R}}=0,
		\end{gathered}
	\end{equation}
	where
	\begin{equation}
		\mathcal{B}_+=\sqrt{\dot{R}^2+1-\frac{2m}{R}-H_0^2R^2}\quad,\quad\mathcal{B}_-=\sqrt{\dot{R}^2+1}.
	\end{equation}
	
	The equation of state derived from Eq.(39) is
	\begin{equation}
		\frac{\dot{\sigma}}{\sigma}=-2\frac{\dot{R}}{R},
	\end{equation}
	 which can be integrated directly as
	\begin{equation}
		\sigma(\tau)=\frac{M_s}{4\pi R^2(\tau)}\quad,\quad M_s=4\pi R^2\sigma=\left(\mathcal{B}_--\mathcal{B}_+\right)R,
	\end{equation}
	where $M_s$, the mass of the shell, remains constant.
	
	Squaring the expression for $M_s$ provides the relationship between the physical quantities and mass of the shell:
	\begin{equation}
		m=M_s\sqrt{{\dot{R}^2+1}}-\frac{M_s^2}{2R}-\frac{1}{2}H^2R^3.
	\end{equation}
	
	We observe that the left-hand side (LHS) of this equation corresponds to the McVittie mass, which is defined as a constant. The terms on the right-hand side (RHS) can be interpreted as the relativistic kinetic energy and binding energy of the shell, subtracting the mass of the interior region when a cosmic background with  $\rho_b=\frac{3H_0^2}{8\pi}$.
	
	For the Schwarzschild case $(H_0=0)$, the quantity $m$ corresponds to the gravitational mass of the shell, as given by the Birkhoff theorem, and the RHS of Eq. (52) represents the conserved energy of the shell.
	
	But for the case with the existence of the cosmological constant ($H_0^2=\frac{\Lambda}{3}$), $M_s\sqrt{{\dot{R}^2+1}}-\frac{M_s^2}{2R}$ is no longer conserved, but varies with the expansion or collapse of the shell. It is also worth noting that if the shell vanishes,  both $m$ and $H_0$ must approach zero, resulting in a global spacetime that is Minkowski.

\subsection{Dust thin shell II: Einstein-Straus model}
\label{sec:G}

The Swiss-cheese model, proposed by Einstein and Straus, represents an idealized universe where local spherical inhomogeneities, such as Schwarzschild regions, are embedded within a homogeneous FLRW background. As two special cases of McVittie spacetime, the relevant physical quantities for this model can be derived using the previously provided formalism.
\\\\

Assuming the interior spacetime is described by the Schwarzschild metric, which is the special case of our model where $H_-=0,m_-=m$, and the exterior spacetime is the spatially flat Friedmann-Robertson-Walker universe with $m_+=0$, we obtain the following relations from Eqs. (42), (43), and (52):
\begin{equation}
		\mathcal{B}_-=\sqrt{\dot{R}^2+1-2m/R}\quad,\quad\mathcal{B}_+=\sqrt{\dot{R}^2+1-H_+^2R^2},
	\end{equation}
		\begin{equation}
		M_s=4\pi R^2\sigma=(\mathcal{B}_--\mathcal{B}_+)R,
	\end{equation}
	and
	\begin{equation}
		m=\frac{1}{2}H^2R^3-\frac{M_s^2}{2R}-M_s\sqrt{\dot{R}^2+1-H_+^2R^2}.
	\end{equation}
	
	Next, if the shell is comoving with the background, i.e.,
	\begin{equation}
		\dot{T}_+=1\quad,\quad \dot{R}=H_+R,
	\end{equation}
	we have
	\begin{equation}
		\mathcal{B}_+=1\quad,\quad K^\tau{}_\tau^+=\frac{\dot{\mathcal{B}}}{\dot{R}}=0\quad,\quad \dot{\Phi}=0,
	\end{equation}
	and the mass of the shell becomes
	\begin{equation}
		M_s=-R + R \sqrt{1- \frac{2 m}{R} +H_+^2 R^2}.
	\end{equation}
	Here, in Eq. (58), the energy conditions are applied to select the physically valid $M_s$
	
	Finally, for smooth matching between the two spacetimes (i.e., $M_s = 0$), the central mass $m$ must satisfy the condition
	\begin{equation}
		m=\frac{1}{2}H_+^2R^3=\frac{4\pi R^3}{3}\rho_+=\mathbf{Const.},
	\end{equation}
	which implies that the background universe must be a dust-filled FLRW model.

\subsection{Cosmic bubbles}
\label{sec:H}

The metric for spatially flat FLRW spacetime is well-known and can be expressed in physical coordinates as
	\begin{equation}
		ds^2=-\left[1-r^2H^2(t)\right]dt^2-2rH(t)dtdr+dr^2+r^2d\Omega^2,
	\end{equation}
	which is a special case of McVittie’s metric in our form by setting $m=0$.
	
	Now, let’s consider the spacetimes inside and outside the bubble, both described by spatially flat FLRW metrics, with arbitrary Hubble functions $H_{\pm}(t_\pm)$. The new functions for these spacetimes are
	\begin{equation}
		\mathcal{B}_\pm=\sqrt{\dot{R}^2+1-H^2_\pm R^2}=\left(1-H^2_\pm R^2\right)\dot{t}_\pm+H_\pm R\dot{R}.
	\end{equation}
	
	Previous studies on bubble dynamics\cite{sakaiJunctionConditionsFriedmannRobertsonWalker1994} have provided the expression for the angular part of the extrinsic curvature of the time-like shell in a spatially flat FLRW spacetime. The metric form is given by
	\begin{equation}
		ds^2=-dt^2+a(t)\left(d\chi^2+\chi^2 d\Omega^2\right).
	\end{equation}
	 When written in Gaussian coordinates, the extrinsic curvature can be concisely expressed as
	\begin{equation}
		K^\theta{}_\theta=\frac{1}{R}\frac{\partial \bar{R}}{\partial n}=\zeta\frac{\gamma}{R}\left(1+vHR\right),
	\end{equation}
	where $\zeta=\text{sign}(\partial\chi/\partial n)$, $\gamma =\frac{\partial t}{\partial\tau}=\frac{1}{\sqrt{1-v^2}}$ is the Lorentz factor, and $v=a(t)\frac{d\chi}{dt}$ is the peculiar velocity of the shell moving relative to the background.

	Under transformation, the peculiar velocity can be expressed in our coordinates as
	\begin{equation}
		v=\frac{dR}{dt}-HR=\frac{1}{\gamma}\dot{R}-HR.
	\end{equation}
	Substituting this expression for $v$ and $\gamma=\dot{T}$ into (63), with $\zeta=+1$, we obtain
	\begin{equation}
		K^\theta{}_\theta=\frac{1}{R}\left[\left(1-H^2R^2\right)\dot{T}+HR\dot{R}\right]=\frac{\mathcal{B}}{R}.
	\end{equation}
	This shows that our expression for the extrinsic curvature is equivalent to the result obtained in Gaussian coordinates.
	
	Now, the mass of the shell can be expressed in terms of the Hubble functions of the two spacetimes:
	\begin{equation}
	\begin{aligned}
			M_s=4\pi R^2\sigma&=R\left(\mathcal{B}_--\mathcal{B}_+\right)\\&=R\left(\sqrt{\dot{R}^2+1-H^2_- R^2}-\sqrt{\dot{R}^2+1-H^2_+ R^2}\right).
\end{aligned}
		\end{equation}
	Following the steps outlined in the previous subsections, we can square the equation and obtain the evolution equation for the mass of the shell:
	\begin{equation}
		\begin{aligned}
			\frac{M_s^2}{2R}-M_s\sqrt{\dot{R}^2+1-H^2_- R^2}=\frac{1}{2}\left(H_-^2-H_+^2\right)R^3=\frac{4\pi R^3}{3}\left(\rho_--\rho_+\right).
		\end{aligned}
	\end{equation}
	
	Squaring $\mathcal{B}_\pm$, then substitute it into (44)
	\begin{equation}
	\begin{aligned}
		\mathcal{B}_\pm
		=\frac{\mathcal{B}_+^2-\mathcal{B}_-^2\pm(\mathcal{B}_+-\mathcal{B}_-)^2}{2(\mathcal{B}_+-\mathcal{B}_-)}=\frac{\rho_+-\rho_-\mp6\pi\sigma^2}{3\sigma}R
		,
	\end{aligned}
			\end{equation}
	which is consistent with the results from\cite{lakeThinSphericalShells1979} and \cite{sakaiBubbleDynamicsSpacetime1993}.
	
	By applying the coordinate transformation used previously, together with the Friedmann equation, we can establish that our result corresponds to the commonly used expression in cosmic bubble studies:
			\begin{equation}
				\begin{gathered}
					\dot{\sigma}+2\frac{\dot{R}}{R}\left(\sigma+\varpi\right)=\frac{1}{4\pi}\left[\frac{H'\dot{T}}{\mathcal{B}}\left(HR- \dot{R} \dot{T}\right)\right]_\pm=[\gamma^2 v(\rho+p)]_\pm,
				\end{gathered}
			\end{equation}
where $v$ is the peculiar velocity of the shell observed in both spacetimes.

\subsection{Cosmological Oppenheimer-Snyder model}

In this subsection, we extend the approach from the previous section, where we derived the dynamics of a cosmic bubble, to study a cosmological version of the Oppenheimer-Snyder model. Specifically, we examine whether our model, which uses the McVittie solution for the exterior and a dust-filled FLRW interior, can reproduce the same equation as (69) under appropriate transformations.

To facilitate our analysis, we use the new definitions of peculiar velocity and Lorentz factor as given by Sakai et al. (2000) in their study of the McVittie spacetime \cite{sakaiPeculiarVelocitiesNonlinear2000}, which are expressed as
\begin{equation}
	v\equiv a(t)\frac{(1+\frac{m}{2 a(t)\bar{r}})^3}{1-\frac{m}{2 a(t)\bar{r}}}\frac{d\bar{r}}{dt} \quad,\quad \gamma\equiv\frac{1-\frac{m}{2 a(t)\bar{r}}}{1+\frac{m}{2 a(t)\bar{r}}}\frac{dt}{d\tau}=\frac{1}{\sqrt{1-v^2}}.
\end{equation}
While these definitions are given in the isotropic coordinate system of the original McVittie metric, it is more convenient to transform to our physical coordinate system where $R$ corresponds directly to the observable physical radius. 
Under transformation, the quantities observed in the McVittie spacetime are
\begin{equation}
	\gamma=\sqrt{1-\frac{2m}{R}}\dot{T};\quad\mathcal{B}\equiv\gamma\left(\sqrt{1-\frac{2m}{R}}+vHR\right);\quad \dot{R}=\gamma\left(\sqrt{1-\frac{2m}{R}}v+HR\right),
\end{equation}
and, with the aid of (2), we obtain
\begin{equation}
	4\pi(\rho+p)|_\Sigma=-\frac{dH(T)}{dT}\frac{1+\frac{m}{2 a(T)\bar{r}}}{1-\frac{m}{2 a(T)\bar{r}}}=-\frac{H'}{\sqrt{1-\frac{2m}{R}}}.
\end{equation}
This shows that the junction condition in the McVittie spacetime can be viewed as an extension of the FLRW case discussed in the previous section, specifically for $m \neq 0$. 

Using the results from (34), (71), and (72), we can demonstrate that our expression for the “energy flux rate” (41) takes the same form when studied in the Gaussian normal coordinate system: 
\begin{equation}
\begin{gathered}
	\dot{\Phi}=\frac{H'\dot{T}}{4\pi\mathcal{B}}\left(HR-\sqrt{1-\frac{2m}{R}} \dot{R} \dot{T}\right)=\frac{(\rho+p)\gamma}{\mathcal{B}}\left(\gamma \dot{R}-HR\right)\\
	=\frac{(\rho+p)\gamma}{\gamma\left(\sqrt{1-\frac{2m}{R}}+vHR\right)}\left[(\gamma^2-1)HR+\gamma^2 v\sqrt{1-\frac{2m}{R}}\right]=\gamma^2 v(\rho+p).
\end{gathered}
\end{equation}
This expression shows that the evolution equation for the shell, as given in equation (39), is consistent with the time component of the conservation identity in the Lanczos equation
\begin{equation}
	\dot{\sigma}+2\frac{\dot{R}}{R}\left(\sigma+\varpi\right)=\frac{1}{4\pi}\left[\frac{H'\dot{T}}{\mathcal{B}}\left(HR-\sqrt{1-\frac{2m}{R}} \dot{R} \dot{T}\right)\right]_\pm=\left[\gamma^2 v(\rho+p)\right]_\pm.
\end{equation}

In the following, we will investigate a modified version of the Oppenheimer-Snyder collapse model, in which the interior is described by a dust-filled FLRW universe, consistent with the original formulation, while the exterior spacetime is replaced by the McVittie solution to account for a more realistic gravitational source in our universe. Here we consider the shell is composed of pressureless matter, co-moving with the interior spacetime. The relevant parameters are
\begin{equation}
	m_-=0,\quad v_-=0,\quad\gamma_-=1,\quad p_-=\varpi=0,
\end{equation}
and the corresponding quantities are given by
\begin{equation}
\begin{gathered}
	\mathcal{B}_-=1,\quad\dot{\Phi}_-=0,\quad \mathcal{B}_+=\gamma_+\left(\sqrt{1-\frac{2m}{R}}+v_+H_+R\right),\\
	 \dot{\sigma}+2\frac{\dot{R}}{R}\sigma=\dot{\Phi}_+=\gamma_+^2 v_+ (\rho+p)\longrightarrow \sigma'+2\frac{R'}{R}\sigma=-\frac{\gamma_+ v_+H'}{4\pi},
	 \end{gathered}
	\end{equation}
	 where the last equation uses the result from Eq. (72), and the prime denotes differentiation with respect to  $t_+$.
    
     To ensure that our model reflects a more realistic universe background, we take the gravitational constant into account, then adopt the Hubble parameter $H_+$ describing an asymptotic $\Lambda$-CDM universe. The analytic solution of the spatially flat $\Lambda$-CDM universe contains both matter and a cosmological constant has the form like\cite{lakeMoreMcVittiesLegacy2011, Hobson2006GeneralRA, Ryden2016IntroductionTC}
     \begin{equation}
         H_+=H(t)=\sqrt{\frac{\Lambda}{3}}\coth(\frac{3}{2}\sqrt{\frac{\Lambda}{3}}t).
     \end{equation}
     This modification introduces a more accurate depiction of the background universe, which allows for a better comparison with current cosmological observations, where dark energy and dark matter play a significant role in the expansion dynamics of the universe.
	
	By applying the initial condition $H_+(t_i)R(t_i)=0.1$, we numerically solved the previous equations (37), (71), and (76). We then explored different internal FLRW models by choosing different initial peculiar velocities. 

    First, we investigated the evolution of the boundary $R(t)$ for the fixed McVittie mass $m$ while varying the initial peculiar velocities. Figures \ref{fig:s1}, \ref{fig:s2}, and \ref{fig:s3} show the results for different initial times $t_i$, which represent approximate conditions corresponding to different epochs in the universe’s history. We observe that the condensed interior will collapse into a black hole when $v_i$ is lower than a critical value, which is influenced by the interior’s Hubble parameter and also depends on the mass parameter $m$. Also notably, the peculiar velocity has a marked influence on the expansion rate of the boundary, particularly during the matter-dominated and dark energy-dominated eras.  We also compare this result with the dust-filled universe model, and find that for the same mass parameter $m$, the critical peculiar velocity for collapse is smaller when the Hubble function includes the cosmological constant. This is due to the expansion effect of dark energy, which counteracts the collapse and lowers the critical velocity threshold. The inclusion of the cosmological constant leads to a suppressed collapse rate, allowing the interior region to stabilize at lower velocities compared to the simpler dust model. 

    In the second set of calculations, we fixed the initial peculiar velocity at $v_i = 0.3$ and varied the mass parameter $m$. This allowed us to explore how the different values of $m$ influence the boundary’s evolution. The results showed that the mass parameter significantly affects expansion dynamics, especially during the transition from matter domination to dark energy domination.

    Moreover, this calculation also provides a verification of the results obtained by Haines et al.\cite{haines1993thin,sakai2000peculiar} in physical coordinates, confirming that negative values of McVittie mass tend to accelerate the boundary’s expansion, consistent with the findings in the literature.
    
    Finally, we computed the time evolution of the Misner-Sharp mass, which is crucial for understanding the gravitational dynamics of the system. According to the previously given metric, the Misner-Sharp mass is the sum of the McVittie mass and the mass of the background cosmic matter enclosed within the interior spherical region. \begin{equation}
        M=m+\frac{1}{2}H_+^2R^2=m+\frac{4\pi\rho_+}{3} R^3
    \end{equation}
    The calculated results, shown in Figure \ref{fig:msmass}, illustrate how the Misner-Sharp mass evolves over time in relation to the boundary’s expansion. 
    
    In the case of collapse ($v_i$ below the critical value), the Misner-Sharp mass decreases monotonically, approaching but slightly exceeding the McVittie mass at the moment of black hole formation. This behavior is intuitive, as the interior region shrinks and the background matter density decreases over time. 
    
    When $v_i$ exceeds the critical value, the Misner-Sharp mass initially declines with time, reaches a turning point, and subsequently begins to rise. This behavior can be understood as follows: during the matter-dominated era, the Hubble parameter decreases rapidly, and the background matter density reduces significantly, but the expansion of the boundary is slow. This results in a temporary decrease in the Misner-Sharp mass. As the universe enters the dark energy-dominated era, the cosmic density stabilizes and the boundary expansion accelerates, causing the Misner-Sharp mass to increase as it sweeps through a larger volume of cosmic matter.
    
    For initial conditions with higher expansion velocities, the Misner-Sharp mass first grows rapidly, then experiences a slight reduction, and ultimately continues to increase with the expansion of the universe.  In this case, the faster expansion of the boundary causes a brief increase in the Misner-Sharp mass, as the boundary sweeps through more cosmic matter. As the peculiar velocity gradually approaches zero over time, the system’s behavior resembles the case with lower initial velocities, where the mass initially decreases and then increases as the expansion accelerates under the influence of dark energy.

\section{Conclusions}
\label{sec:5}
In this paper, we present the matching conditions for concentric McVittie spacetimes, with a detailed focus on deriving the extrinsic curvature and surface stress-energy tensors under various configurations. By transforming the McVittie metric into a physical coordinate system, we clarify its behavior in different regimes and construct a theoretical model in which McVittie spacetimes describe both the interior and exterior.

 We demonstrate that our results are consistent with all previous remarkable cases, including Schwarzschild, FLRW, and Einstein-Straus solutions, highlighting the robustness of the thin-shell framework. Furthermore, the introduction of the function $\mathcal{B}$ provides a simplified yet comprehensive representation of the surface quantities, offering new perspectives for analyzing thin-shell dynamics.

 We analyze the junction condition in the context of McVittie spacetimes, derive the expression for the energy flux in McVittie parameters, and extend the study to a modified Oppenheimer-Snyder model. By numerically solving the governing equations for different initial conditions, we explore the evolution of a thin shell connecting two McVittie spacetimes. The numerical results reveal how the mass parameter $m$ and the initial peculiar velocity $v_i$ influence the boundary’s evolution. Specifically, we observe that the interplay between $m$, $v_i$, and the interior Hubble parameter determines whether the boundary expands indefinitely or collapses into a black hole. Notably, the effect of negative McVittie mass was explored, showing distinct behaviors in both expanding and collapsing scenarios. Additionally, the study of Misner-Sharp mass reveals crucial insights into the gravitational dynamics of the system. These findings provide insights into the dynamics of thin shells in cosmological settings, offering a broader perspective on junction conditions and their applications to generalized McVittie spacetimes. 
 
 In conclusion, this study enhances our understanding of the dynamics of dynamic thin shells and their role in cosmological and astrophysical contexts. it offers important implications for future research in generalized McVittie spacetimes.


\bibliography{references}

\begin{figure}[p]
  \centering
  \begin{subfigure}[t]{0.8\textwidth}
    \includegraphics[width=\linewidth]{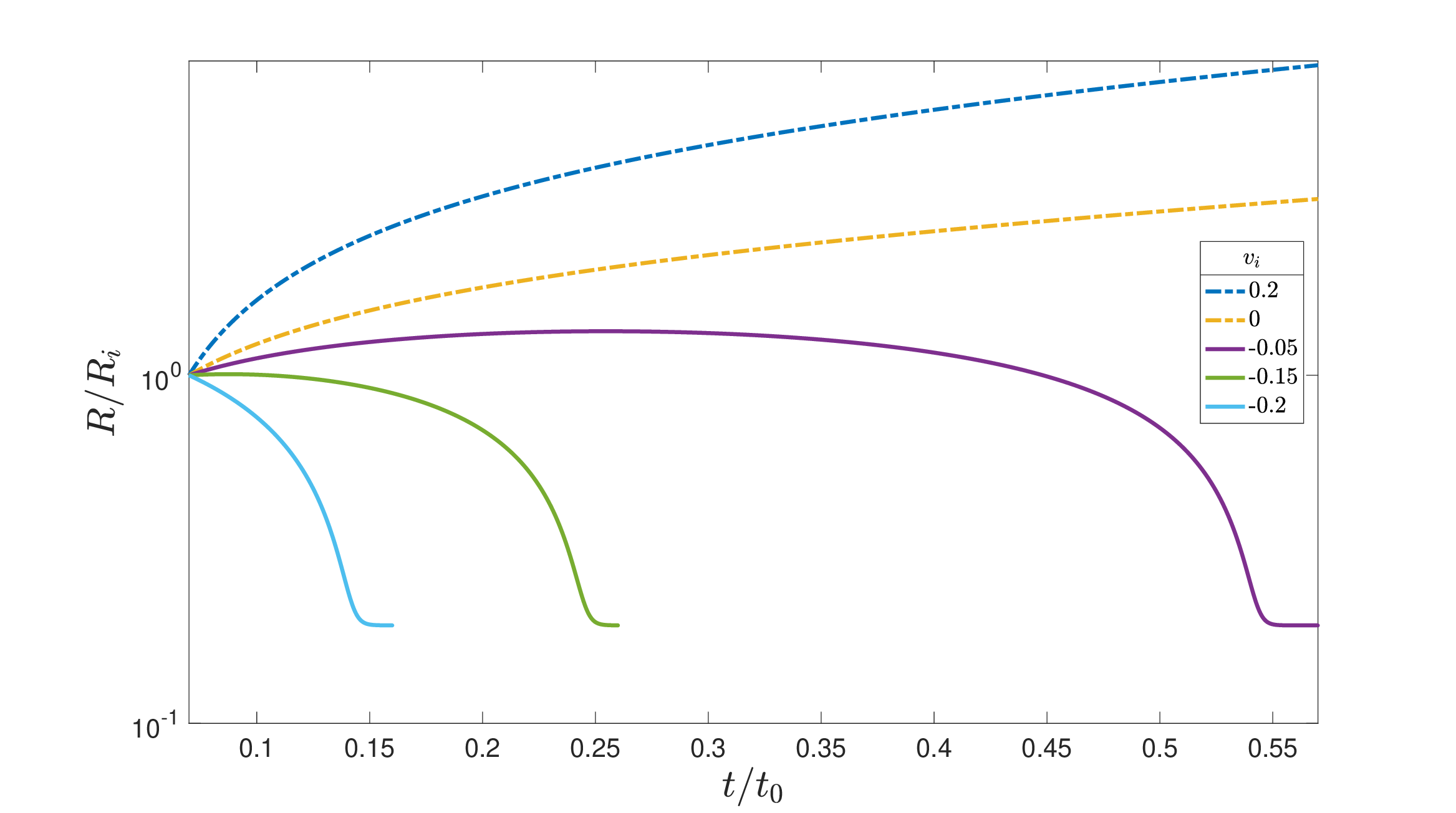}
    \caption{The initial time $t_i = 0.07 t_0$, approximately represents the early universe, where matter dominates and dark energy has a negligible effect.}
    \label{fig:s1}
  \end{subfigure}
  
  \vspace{1cm}
  \begin{subfigure}[t]{0.8\textwidth}
    \includegraphics[width=\linewidth]{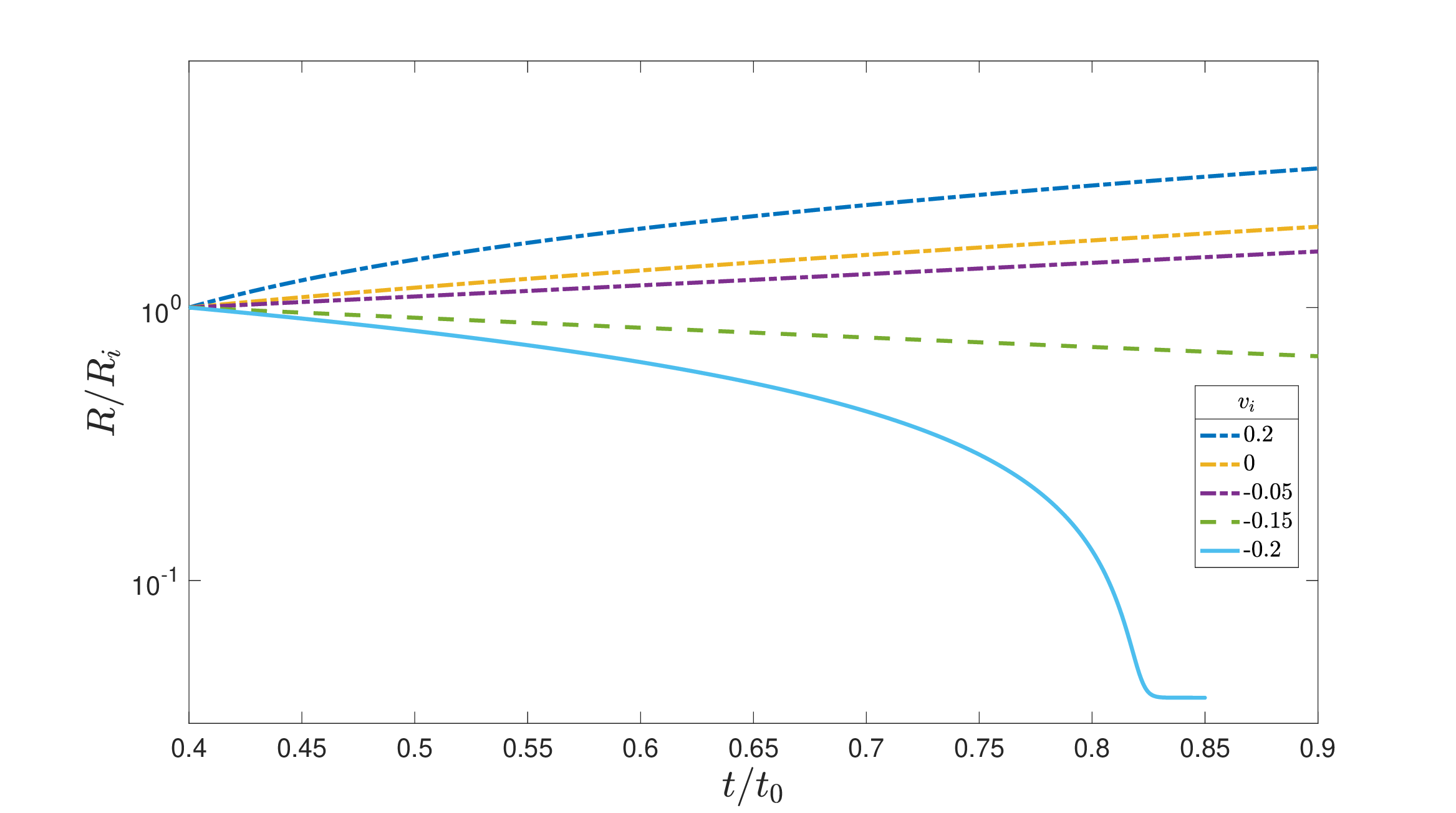}
    \caption{The initial time $t_i = 0.4 t_0$, corresponds to a time when both matter and dark energy contribute significantly to the expansion of the universe. }
    \label{fig:s2}
  \end{subfigure}
  
  \caption{Time evolution of the normalized boundary radius $R/R_i$ for different initial peculiar velocities $v_i$ with fixed mass parameter $m = 0.01$.The results are shown for three different initial times $t_i$, each corresponding to distinct epochs in the universe’s history(continued on next page)}
  \label{fig:main}
\end{figure}

\begin{figure}[p]
  \ContinuedFloat 
  \centering
  \begin{subfigure}[t]{0.8\textwidth}
    \includegraphics[width=\linewidth]{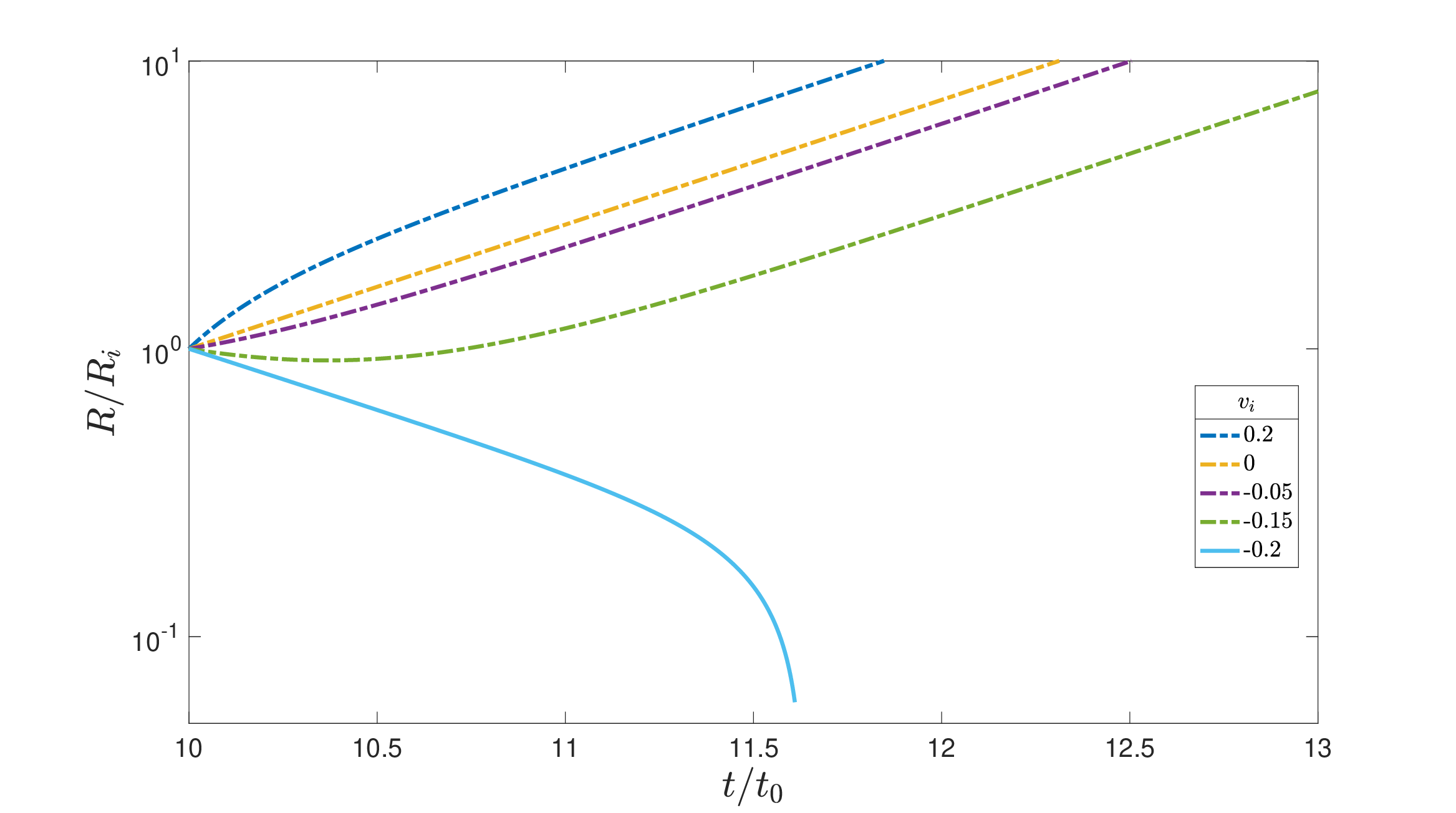}
    \caption{The initial time $t_i = 0.4 t_0$, represents the future epoch where dark energy dominates the expansion.}
    \label{fig:s3}
  \end{subfigure}
  
  \caption[]{Time evolution of the normalized boundary radius $R/R_i$ for different initial peculiar velocities $v_i$ with fixed mass parameter $m = 0.01$.The results are shown for three different initial times $t_i$, each corresponding to distinct epochs in the universe’s history (cont.)} 
\end{figure}

\begin{figure}[htbp]
    \centering
    \includegraphics[width=0.8\textwidth]{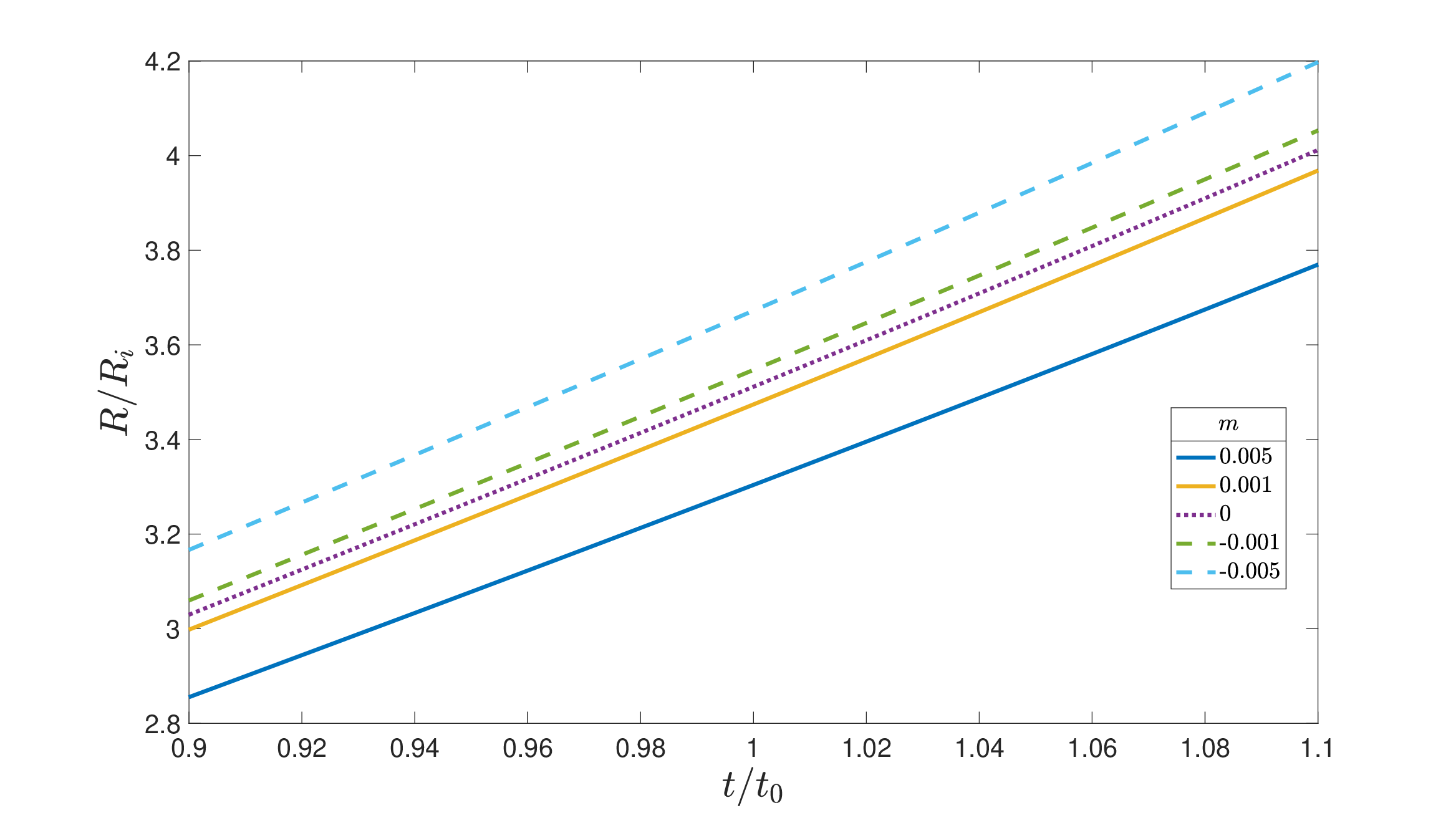}
    \caption{The effect of different values of $m$ on the boundary evolution, with the initial peculiar velocity $v_i = 0.3$.}
    \label{fig:v03m}
\end{figure}

\begin{figure}[htbp]
    \centering
    \includegraphics[width=0.8\textwidth]{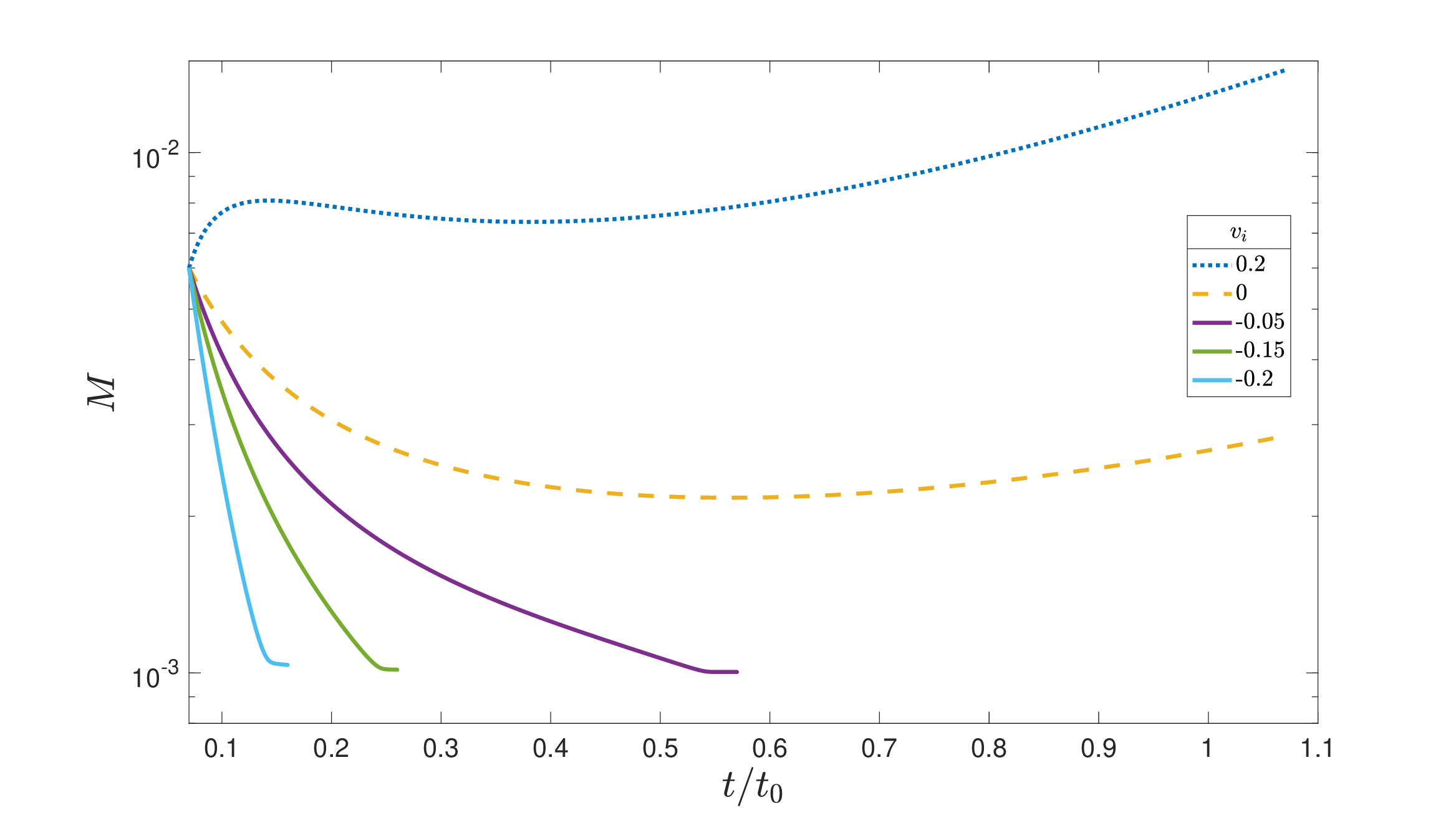}
    \caption{Evolution of the gravitational mass $M$ with time for different $v_i$. The mass parameter is set to $m = 0.01$, as shown in the Figure \ref{fig:main}.}
    \label{fig:msmass}
\end{figure}

\end{document}